\documentclass[preprint]{aastex631}

\makeatletter
\renewcommand{\frontmatter@title@above}{}
\makeatother

\renewcommand*{\thefootnote}{\fnsymbol{footnote}}

\begin{document}

\begin{figure}
\centering
\includegraphics[width=3.5cm]{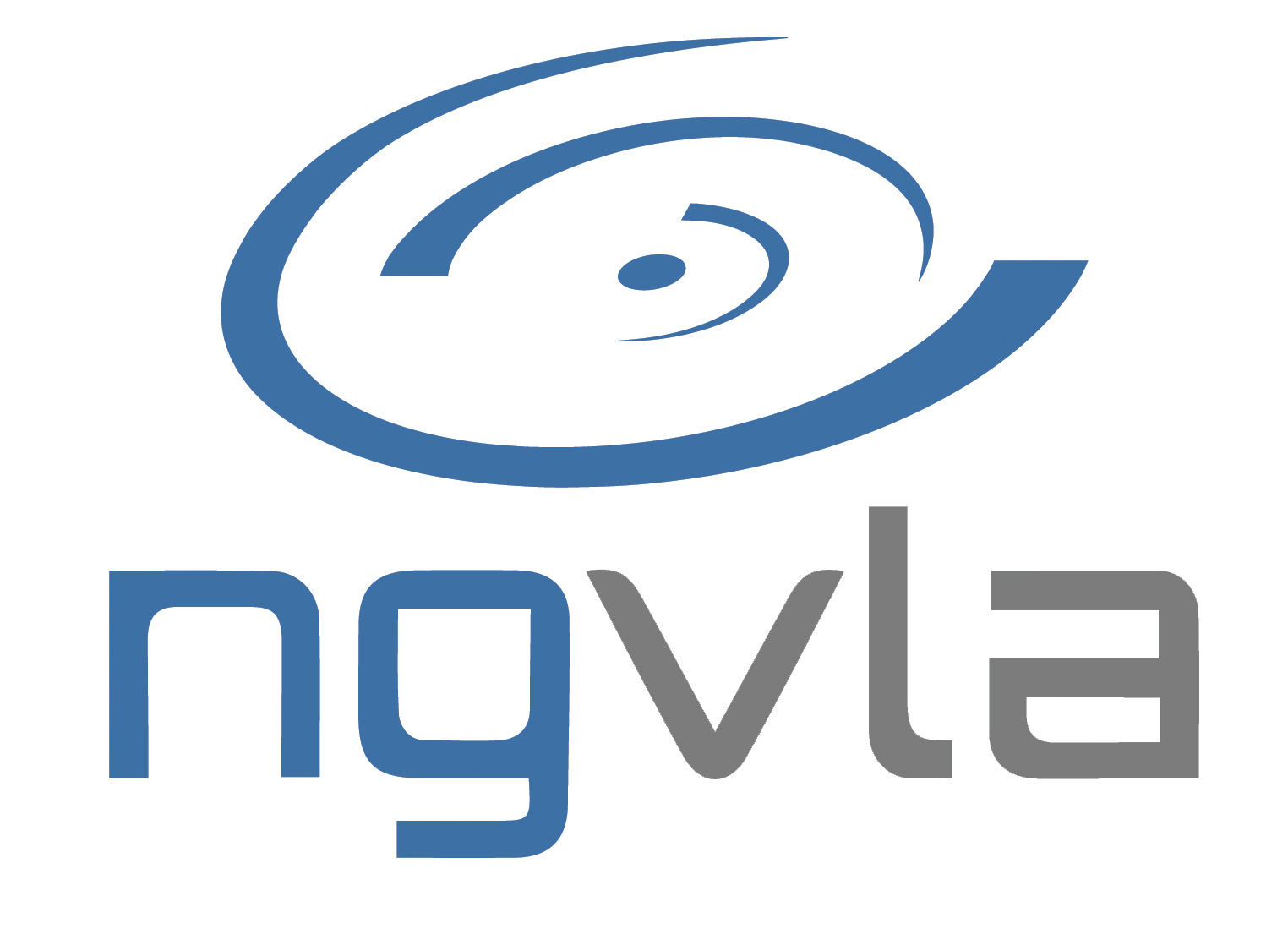}
{\center\Large\bf ngVLA Memo No. 138 }
\end{figure}

\title{An Assessment of Water Vapor Radiometry Landscape in ngVLA Context }
\author{T. K. Sridharan}
\affiliation{National Radio Astronomy Observatory, 520 Edgemont Rd. Charlottesville, VA 22903, USA}



\pagestyle{plain} 
\pagenumbering{arabic}

\pagestyle{plain} 
\pagenumbering{arabic}

\begin{abstract}

The ngVLA intends to utilize Water Vapor Radiometry (WVR) as an approach to lower calibration overheads for tropospheric phase fluctuation correction in its high frequency bands (5 \& 6; 30.5-116 GHz) to levels below those feasible with fast switched reference gain calibration. As a stepping stone to an eventual ngVLA WVR system, the JVLA/ngVLA teams pursued an effort to develop and install a set of MMIC based compact water vapor radiometers (CWVR) and carry out on-sky testing. Most recent tests found that the CWVRs installed on the VLA antennas under this effort were non-functional in tracking visibility phases. As part of the program to develop a WVR system for the ngVLA, this memo summarizes the current landscape of water vapor radiometry experience, the situation with the CWVRs, how they relate to the ngVLA WVR requirements, and the work needed to operationalize the CWVRs as useful test systems and to make broader progress. We currently lack statistically relevant WVR correction data to understand their expected performance for the ngVLA sites. We also discuss parallel additional approaches alternative and/or complementary to the CWVRs. A path involving tests to establish the functionality, stability, sensitivity, and switched power cal/data synchronism of the CWVR system in the lab and on antenna, followed by on-sky tests, and a parallel effort to assess the utility of autocorrelation outputs of the VLA K-band receivers for delay estimation is recommended. Potential paths going beyond traditional excess path estimation methods which may be necessary to meet ngVLA requirements and goals are touched upon.     

\end{abstract}

\section{Introduction \& Motivation} 

The prospective technique adopted in the baseline ngVLA design to improve the calibration efficiency for tropospheric delay fluctuation correction on short time scales for  high frequency observations, primarily in bands 5 \& 6, is water vapor radiometry (WVR) \citep{Clarke2015WVRngVLAMEM10, 2020ngVLAMEM061, Towne2020WVRngVLAMEM79, Butler2020ngVLAMEM074, 2024ngVLASysConDesRep, 2024ngVLACalConcept, Sridharan2025ngVLAEmemo21Band-4WVR, Asaki2025ngVLAMEM135}; see earlier work by \cite{1999RaSc...34..817C} and \cite{1999VLAMEMO177} for extensive considerations and modeling in the context of the VLA and \cite{ChandlerEVLA_Memo74_CWVR_2004,ChandlerEVLA_Memo73_WVR_2004} and \cite{2018arXiv180701690G} for experimental VLA prototypes. The technique, as applied to the ngVLA sites, detects the 22 GHz water vapor line emission from the troposphere and uses its parameters to estimate excess delay on short time scales for real time and offline corrections (\citealt{1999VLAMEMO177} includes a good overview and further references to previous work). 

The ngVLA WVR must deliver performance fulfilling the ngVLA science and system requirements which call for image dynamic ranges of 32 and 28 dB in Bands 5 \& 6 and a calibration efficiency goal  (fraction of time spent on source)  of 90\% \citep{2021ngVLASysReq}. The resulting overall implementation agnostic requirements for the measurement accuracy and timescales of the ngVLA WVR system have been derived in \cite{Asaki2025ngVLAMEM135}. As to system implementation concept and design, two paths have been proposed and compared: a standalone, small dish WVR \citep{2022WVRDes, 2022WVRTechReq} and an alternative Band-4 WVR concept \citep{Sridharan2025ngVLAEmemo21Band-4WVR} which uses the main ngVLA antennas and Band-4 science receivers with important technical and programmatic advantages. Fast-switching to a nearby reference calibrator is the standard approach to tropospheric delay fluctuation correction which provides the default backup strategy for ngVLA, with a calibration efficiency of $\sim$ 50\%. This immediately leads to a high level requirement that any WVR system must deliver  success rates in improving residual phase fluctuations substantially better than $\sim$ 50\% for its pursuit to be meaningful. 

\section{WVR System Outline}

A WVR system consists of a water vapor sensing hardware subsystem, in turn comprising a $\sim$~K-band frontend covering the water line and a backend to construct coarse spectra, nominally one unit at each antenna, and an algorithm/software subsystem. The algorithm uses the sensed spectral data to estimate tropospheric delay corrections to be applied to concurrently obtained interferometric science data or in the phasing system. The WVR requirements derived in \cite{Asaki2025ngVLAMEM135} are at the level of the full overall WVR system. The work to further break down these requirements into WVR hardware requirements (frequency coverage, number and widths of the spectral channels, sensitivity, and stability) and path length prediction algorithm accuracy requirements through tropospheric water line modeling, and the development of such an algorithm are in progress (Massingill et al). Empirical and physics guided empirical methods are also possible and are also being pursued (Svoboda et al). These efforts build on previous work, most recent of which are in \cite{Butler2020ngVLAMEM074} and \cite{Towne2020WVRngVLAMEM79}, which also include references to prior research.

There are three elements to the requirements: (1) the raw thermal sensitivity to adequately detect the water vapor emission; this is not expected to be difficult to achieve e.g. \cite{Sridharan2025ngVLAEmemo21Band-4WVR} (2) the stability of the system; this can be difficult depending on the applicable time scales and the specific implementation - standalone vs Band-4 - with significantly less stringent requirement for the latter \citep{Sridharan2025ngVLAEmemo21Band-4WVR} and (3) the accuracy of the delay estimation algorithm, which may be the weakest link.

The CWVR was intended to be a pathfinder to assess the viability of WVR corrections for the EVLA, and more recently the ngVLA, to gain experience, and to provide a tool for the acquisition and analysis of statistically meaningful data specific to the VLA/ngVLA sites.  

\section{Past Experience}

\begin{figure}[h!]
\begin{center}   

\includegraphics[width=0.7\linewidth]{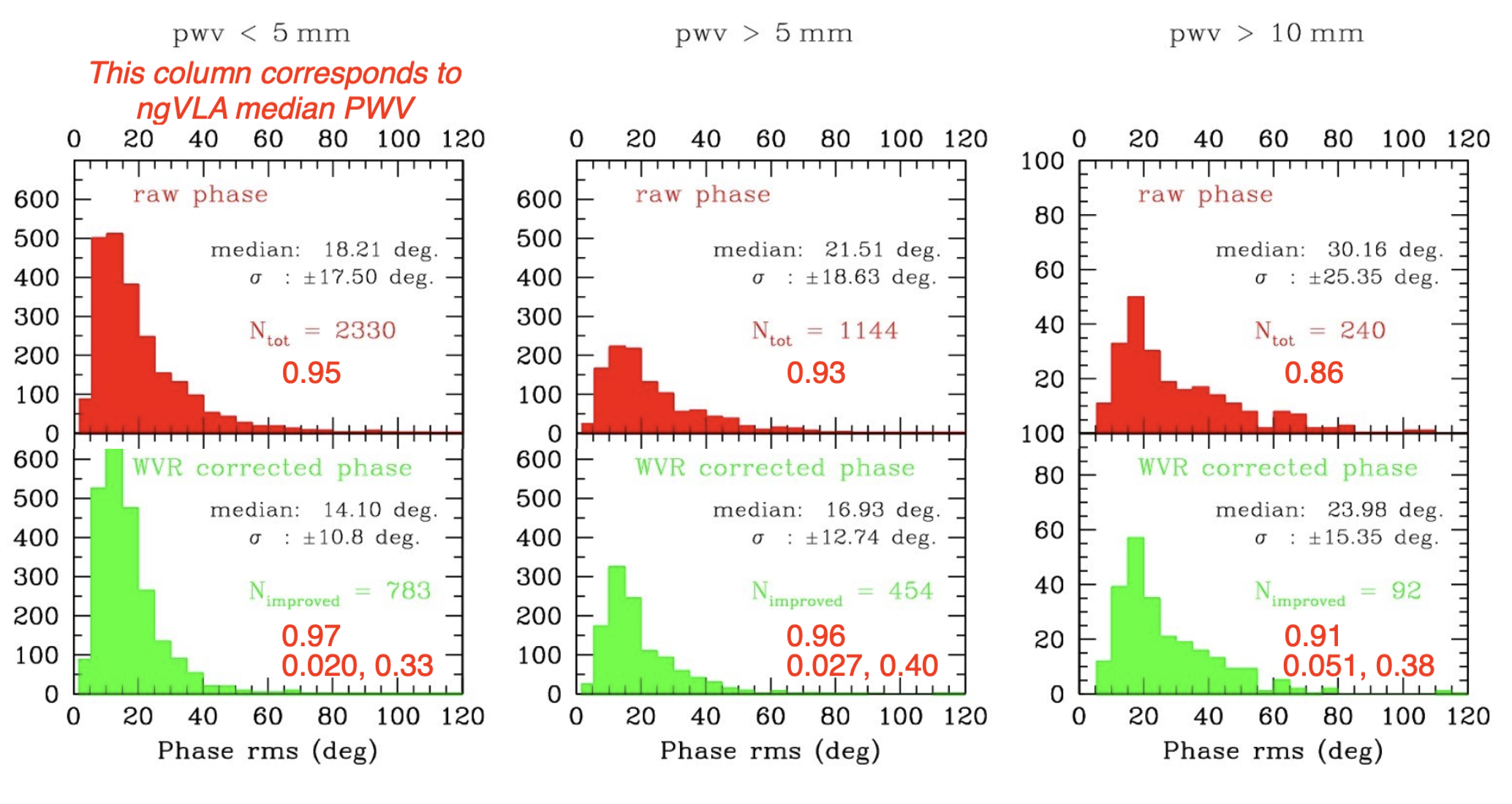}
\caption{\small: NOEMA WVR performance in the 3-mm band. The distribution of raw uncorrected phases (top row) and that of WVR corrected phases (bottom row) are shown for different PWV conditions. The computed coherence factor per baseline for the median phase  error is listed in red in each panel. The bottom row includes the coherence factor after correction, the achieved improvement in the coherence factor, and the fraction of data that gained improvement. The left column corresponds to PWV conditions similar to those applicable to the ngVLA sites for observations in bands 5 \& 6 (adapted from \citealt{Bremer2016_NOEMA_WVR_IRAM_School_talk}; text in red are our additions).}
\vspace{-5mm}
\label{Fig: NOEMA}
\end{center}
\end{figure} 

WVR systems in an interferometric phase correction context are currently operational at NOEMA, ATCA, and ALMA. We will not consider the highly successful ALMA system further as it uses the 183 GHz water vapor line which is appropriate for the very low water vapor levels at the ALMA site and is therefore not directly applicable to the ngVLA sites. The NOEMA and ATCA systems use the 22 GHz line and the sites have water vapor levels comparable to the VLA site which is taken to be representative of the ngVLA sites. However, there is very little long term statistical analysis of the performance of these systems available in the literature. The most extensive we are aware of is from NOEMA, as presented at the 2016 IRAM Interferometry School \citep{Bremer2016_NOEMA_WVR_IRAM_School_talk}, which we will use as indicative of the performance of the NOEMA WVR system. 

Figure~\ref{Fig: NOEMA} shows the results of phase correction in the 3-mm band at NOEMA using their currently operational generation of 3-channel WVR (a subset of the antennas have a newer 14 channel system). While we do not have many details about the data, as shown in the figure, they provide an important takeaway  that WVR corrections deliver improvements $\sim$ 40\% of the time at best. For the PWV conditions applicable to ngVLA Band 5 \& 6 observations (left most column), the success rate is $\sim$ 33\%. While the $<$ 5 mm PWV conditions are applicable to ngVLA high frequency observations, the data correspond to very good phase stability conditions as can be seen from the low median phase rms and may thus not be representative, this is the most extensive statistics available to our knowledge. However, the rms after correction are in a range approaching the ngVLA requirements (section 4, Table 1) and therefore relevant as useful indicators of performance.   

Coming to experimental test systems, limited information for the VLA sites is available from two NRAO efforts. \cite{ChandlerEVLA_Memo73_WVR_2004} present results from $\sim$ hour-long single baseline on-sky test observations using a 3-channel system they developed, spanning three different baseline lengths and different weather conditions, with derived corrections applied to calibrator data at 22~GHz and 43~GHz. The residuals were 570~$\mu$m of path length (1.9~ps), improving from 1270~$\mu$m (4~ps) for an 800~m baseline under clear skies and 1030~$\mu$m of path length (3.4~ps), improving from 2660~$\mu$m (9~ps) for a 2.5 km baseline under forming cumulus, both for corrections applied to 22~GHz data, and 500~$\mu$m of path length (1.7~ps), improving from 2085~$\mu$m (7~ps) for a 2.5~km baseline under clear skies for corrections applied to 43~GHz data. These numbers are over 10~min time scales and were obtained by converting the phase residuals reported in \cite{ChandlerEVLA_Memo73_WVR_2004} to frequency independent quantities - path lengths and delays. While the results clearly show that the corrections provided substantial improvements for these observations, they cannot be taken as statistically representative and are insufficient to assess expected success rates for improvement. 

The next generation of the NRAO effort is represented by the CWVR, which grew out of the first generation discussed above, with a 5-channel design using monolithic microwave integrated circuit (MMIC) filters resulting in a compact system affording better thermal control \citep{ChandlerEVLA_Memo74_CWVR_2004, 2018arXiv180701690G, Towne2020WVRngVLAMEM79, Butler2021WVRngVLAMEM91}. The best reported test results  for this system \citep{Towne2020WVRngVLAMEM79} showed $\sim$~2~ps residuals ($\sim$~300~$\mu$m path error). This is again based on $\sim$~1 hr of data and cannot be used to assess expected success rates for improvement. Importantly, as already noted, it has not been possible to reproduce  these results (Butler, private communication), with the most recent tests showing that the CWVRs are non-functional as radiometers and not successful in tracking visibility phases \citep{Butler2021WVRngVLAMEM91}.

\section{Comparison with Requirements}
Even though the existing results are either not extensive enough to be statistically representative or are for a different site, it is useful to compare them to the ngVLA requirements derived in \cite{Asaki2025ngVLAMEM135} which can be summarized as: $\sim$ 250~$\mu$m ($\sim$~0.8~ps) and $\sim$ 170~$\mu$m ($\sim$~0.5~ps) for 80\% coherence on 1-2~s and 3-4~s time scales and $\sim$ 170~$\mu$m ($\sim$~0.5~ps) and $\sim$ 150~$\mu$m ($\sim$~0.5~ps) on 1-2~s and 3-4~s time scales for 90\% coherence. Table~\ref{tbl01} presents this comparison.

\begin{deluxetable*}{c|c|c|c|c}[h!]
\tablenum{1}
\tablecaption{
A comparison of the ngVLA WVR requirements and results from operational/experimental systems 
\label{tbl01}}

\tablehead{
 System & time scale & path error  & delay error & Comment
}
\startdata 
\hline
ngVLA Req \citep{Asaki2025ngVLAMEM135} & 1~s    &   264-173~$\mu$m  &  0.9-0.6 ps & for 80-90\% coherence\\
& 2~s    &   246-167~$\mu$m  &  0.8-0.6 ps  & \\
& 4~s    &   141~$\mu$m  &  0.5 ps & \\
\hline
CWVR \citep{Towne2020WVRngVLAMEM79} & $\sim1$~s    &   300~$\mu$m  &  2 ps & 1-hr data;10 km baseline \\
Most recent on-sky \citep{Butler2021WVRngVLAMEM91}& - & - & - & non-functional as WVR\\
\hline
NOEMA \citep{Bremer2016_NOEMA_WVR_IRAM_School_talk} & $\sim$~min & 140~$\mu$m & 0.5 ps & Median; $\sim$~33\% success rate \\
\hline
VLA-WVR \citep{ChandlerEVLA_Memo73_WVR_2004} & 10 min    &   570~$\mu$m  &  1.9 ps & $\sim$ 1 hr long data   \\
(clear skies) & & & & 800 m baseline \\
&     &   500~$\mu$m  &  1.7 ps & $\sim$ 1 hr long data \\
& & & & 2.5 km baseline \\
\hline
\enddata
\end{deluxetable*}
\vspace{-20mm}

\section{CWVR Status}
Figure 2 presents the comparison in a different way along with the lab and on-sky performance status of the CWVR. It is an adaptation of Fig 15 of \cite{2018arXiv180701690G}, where the results of Allan variance stability tests of the prototype CWVR were presented. A $T_{sys}$ of 100 K was assumed to scale the stability to obtain a temperature scale needed to translate the ngVLA requirements, the NOEMA performance, the \cite{Towne2019Precision, Towne2020WVRngVLAMEM79} CWVR performance, and the \cite{Butler2021WVRngVLAMEM91} results. This value of $T_{sys}$ is representative of the \cite{Butler2021WVRngVLAMEM91} measurements and is in the middle of the range applicable to ngVLA, lower for the Band-4 WVR and higher for the standalone WVR. While the exact locations of these systems in the plot can move up or down based on the specific $T_{sys}$ applicable (by 0.3 for a factor 2), the figure is indicative of the performance landscape based on lab and on-sky tests of different realized systems.

\begin{figure}[h!]
\begin{center}  
\includegraphics[width=0.8\linewidth]{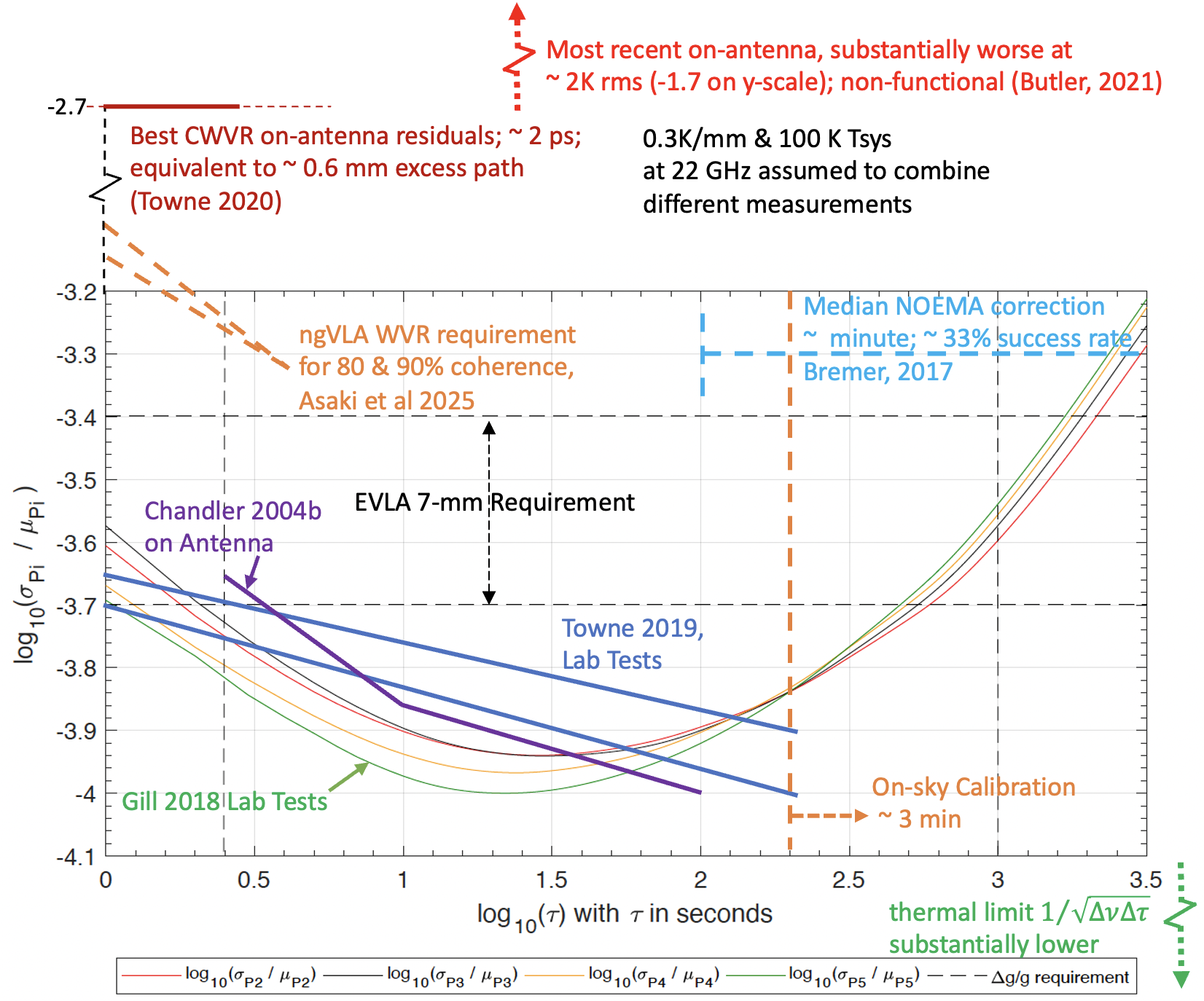}
\caption{\small: A graphical summary of the performance of operational and experimental test WVR systems and ngVLA WVR requirements. The comparison is shown on a lab stability test plot  of the prototype CWVR (Allan Standard Deviation; $\sigma$ normalized by mean power $\mu$ for WVR channel $i$), adapted from \cite{2018arXiv180701690G}. A $T_{sys}$ of 100 K and a 22 GHz water line strength to excess path length due to water vapor ratio of 0.3 K/mm have been assumed to plot the NOEMA performance and the CWVR performance. See text for more discussion.} 
\label{Fig: WVR_Landscape}
\end{center}
\end{figure}

Clearly, the best achieved results of 1.7 - 2 ps compared to the 0.5 - 0.8 ps needed (on different timescales) and the very limited data available from past efforts for ngVLA-representative sites point to the need for improvement and the acquisition of more and directly relevant data. This was partly the purpose of the CWVR whose debugging and operationalization must therefore take high priority followed by further development and tests to assess if the ngVLA requirements can be met and to quantify the expected success rates for realizing improvements at the ngVLA sites. The lack of adequate staffing has prevented progress towards these goals. Attempts to recruit past NRAO staff with WVR/reciever experience, to demonstrate the CWVR performance and/or to debug the system were unsuccessful.


Apart from the shortcomings noted in \cite{Butler2021WVRngVLAMEM91}, initial explorations also indicated issues with the separation of ON and OFF states of the switched power noise source in the response of the CWVR system due to its rise time and evidence of strong correlation between ON and OFF data with the noise source switched off, contrary to the expectation that they should be independent (private communications, K. Sowinski, B. Butler). 
The slow time constant may also have detrimental implications for the effective bandwidth, depending on how it arises. 


\section{Path Forward}

Ability to acquire statistically representative data for the ngVLA sites (with the VLA site as proxy) is key to advancing WVR application to ngVLA. 
Central to this is (1) verifying and reproducing \cite{Towne2020WVRngVLAMEM79} performance, which has not been possible \citep{Butler2021WVRngVLAMEM91}, pointing to possible fundamental issues with the system, and (2) understanding, and if possible, eliminating the reasons for this and establishing the current baseline performance. It is hard to reconcile the two - a 2~ps performance in Sep 2019 and its lack of repeatability over Jan 2020 - Jun 2021 with the same system, only a year later. Adequate explanations of the mismatch have not been found. However, it should be noted that the lab tests show that the system would comfortably meet the requirements and the \cite{Towne2020WVRngVLAMEM79} on-sky performance results, if reproducible,  are at a very interesting level for meaningful tests relevant to ngVLA (Fig 2). Therefore efforts to understand the test results mismatch and to debug the system are necessary. Accordingly, in addition to the points noted in \cite{Butler2021WVRngVLAMEM91}, the following steps are suggested.

\subsection{CWVR}
\begin{enumerate}
\item Conduct tests on a lab CWVR unit to demonstrate and quantify functionality, stability, and sensitivity.
\item Conduct lab tests to verify ON-OFF switching signal synchronization with the output switched power patterns and measure and compare rise times for the switching signals and the outputs.
\item Conduct on-antenna sensitivity and stability tests - the CWVR deployment and commissioning processes seem to have skipped this key step and proceeded directly to a short on-sky test whose results have turned out to be non-reproducible. This will establish that the lab performance is reliably reproduced after installation on the antenna without excessive degradation.   

Some of these steps are currently being undertaken by Kyle Massingill, although impacted by lab availability constraints arising from the recent lab flooding incident.

\end{enumerate}

\subsection{Parallel Efforts}
\begin{enumerate}
    \item Pursue VLA-K band switched power autocorrelation based tests with focus on standalone or physics guided empirical methods. Brian Svoboda is leading efforts towards this goal. 
    \item Evaluate the use of stratospheric ozone line emission for measuring water vapor column through its absorption in the troposphere. This approach was proposed by \cite{PaineOzone_2008} for sub-mm wavelength phase corrections using the 230 GHz ozone line. Weaker ozone lines are present in the 3-mm band accessible to the ngVLA and can in principle be utilized. Preliminary explorations showed that the lines may be weak and of limited value, but actual observations by the earth science community measure a factor of $\sim$ few stronger lines (Butler and others; Calibration WG discussions). 
    
While these approaches may not deliver adequate levels of standalone sensitivities to the path length fluctuations, they offer complementary methods. 

\end{enumerate}

As noted before, the ability to convert the WVR data to delay estimates may be the most challenging part of attaining the required WVR accuracy and success rate of improvement if approached {\it ab initio}, as they involve atmospheric physics and radiative transfer processes not well understood at the required accuracy levels. The ozone measurement uses absorption and is therefore a more direct probe of the absorbing column as opposed to the the emission measurements WVRs rely on, where the details of vertical temperature and water vapor profiles are folded in. The empirical methods do not attempt to fully understand the atmospheric radiative transfer.  

We recognize that to reach the delay estimation accuracies and sucess rates needed for the ngVLA, we need to step beyond the traditional methods. The algorithms envisioned (Massingill et al, Svoboda et al) would do so by attempting to place the delay estimation problem in the local neighborhood of the actual applicable location in the parameter space of the atmospheric state using as much ancillary data as possible such as surface weather, global circulation model reanalysis data such as MERRA-2 and ERA5, autocorrelation data from a relevant science receiver, ozone measurements, and calibrator observations on time and/or spatial scales coarser than required for the estimation. The algorithm would seek to derive just a single parameter on finer scales - the excess path length (due to water vapor) - using water vapor data with high spatial and time granularity acquired by the WVR (per-antenna and seconds) in this local neighborhood. This approach appears suited to possible machine learning methods which is beginning to be explored. As the data from the RPG tipping radiometer installed at the VLA site can provide important inputs through vertical temperature and water vapor profiles as we develop these methods, it should be brought back into stable operation. 

In summary, a combination of the above methods synthesized synergistically offers our best path to achieving the required tropospheric delay fluctuation correction accuracies and improvement success rates. Importantly, the path outlined above provides the on-sky verification tools necessary to experiment, develop, and validate delay retrieval algorithms.


\section*{Acknowledgement}

We thank Kyle Massingill, Bryan Butler, and Claire Chandler for helpful comments which led to improvements and Chris Carilli, Rob Selina, Jeff Mangum, and Eric Murphy for reading the draft manuscript. 


\begin{figure}[h!]
\begin{center}   
\includegraphics[width=6.2in]{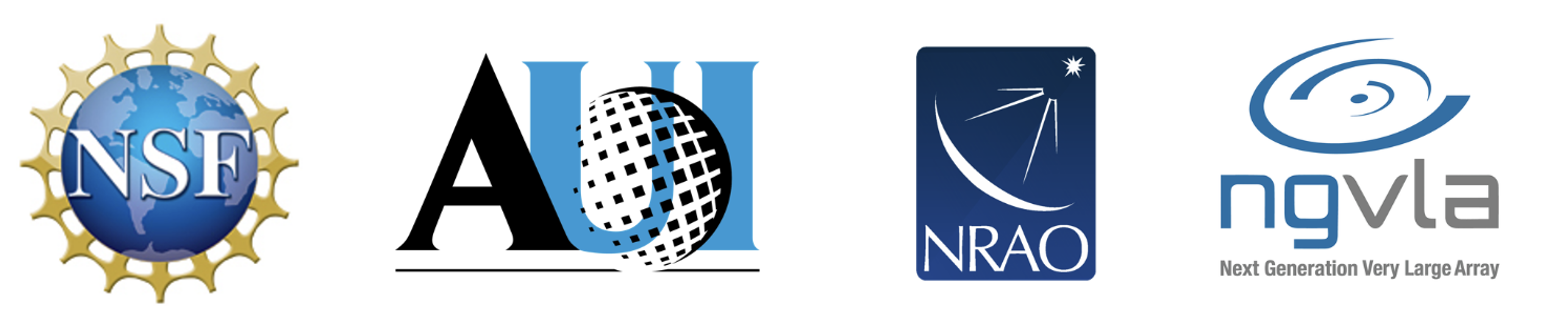}

\end{center}
\end{figure} 

{\it The National Radio Astronomy Observatory and Green Bank Observatory are facilities of the U.S. National Science Foundation operated under cooperative agreement by Associated Universities, Inc. This work was supported by awards AST-2034328 (MSIP Prototype Antenna) and AST-2334267 (ngVLA Design Activities); NRAO related activities are funded under award AST-1647378 (NRAO Operations/Development).}

\bibliography{PASPsample631}{}
\bibliographystyle{aasjournal}

\end{document}